\documentclass{article}

\PassOptionsToPackage{numbers,compress}{natbib}
\usepackage[preprint]{neurips_2026}

\usepackage[utf8]{inputenc}
\usepackage[T1]{fontenc}
\usepackage{hyperref}
\usepackage{url}
\usepackage{booktabs}
\usepackage{makecell}
\usepackage{amsfonts,amsmath,amssymb}
\usepackage{textcomp}
\usepackage{graphicx}
\usepackage{microtype}
\usepackage{xcolor}
\usepackage{placeins}
\usepackage{xurl}
\usepackage{tikz}
\usetikzlibrary{arrows.meta,positioning,fit,backgrounds}

\title{Adaptive KV Retention for LLM Agents at Human-Approval Timescales}

\author{
  Minseo Choi \\
  Johns Hopkins University\\
  \texttt{mchoi46@jhu.edu} \\
  \And
  Ananya Joshi \\
  Johns Hopkins University \\
  \texttt{aajoshi@jhu.edu}
}

\begin{document}

\maketitle

\begin{abstract}
Unlike the seconds-scale tool-call pauses targeted by prior agent-serving systems, agentic LLM requests can be suspended for minutes or hours while waiting for human approval. We study how suspension and resumption affect GPU serving performance and develop a retention policy that balances active-serving capacity against future recomputation under uncertain approval waits. The central tension is severe because retaining suspended KV preserves fast resume but can consume enough GPU capacity to reduce active-serving goodput by 41\%, while evicting it avoids that residency cost at the expense of nearly $10\times$ higher resume latency when the request returns. We develop a tiered retention controller around \textit{GPU opportunity cost}, which expresses the serving capacity consumed by preserving or reconstructing a suspended request's KV state in a common GPU-time cost. Within host memory, the controller selects between indefinite retention and load-indexed expiration using calibration wait samples, without requiring per-request wait prediction. On human-scale approval workloads, our controller improves active-request goodput by 23--51\% over the vLLM baselines, 22--29\% over MORI, and 41--52\% over Continuum.
\end{abstract}

\section{Introduction}
\label{sec:intro}

Agentic LLM pipelines increasingly pause before consequential actions such as sending email, changing files, or running user commands. Coding assistants already wait for human approval when they request permission for such actions \citep{claudecode2025, openaicodexapp2026, githubcopilot2026}, and orchestration frameworks expose human approval as an interrupt primitive \citep{langgraph2026, wu2024autogen}. Still, existing agent-serving systems, like MORI and Continuum, mainly study tool-call pauses on the scale of seconds \citep{continuum2025,mori2026}.\footnote{Continuum explicitly targets tool calls as short as $\leq 2$\,s, and MORI's coding-agent traces report median and 90\% tool-call durations of 1.10\,s and 2.03\,s, respectively, while listing human approval as an example of a longer idle phase.} Human approvals, however, can last minutes to hours. This longer idle regime creates a retain-or-release tradeoff similar to serverless keep-alive policies, which preserve idle functions to avoid cold starts while limiting their resource footprint \citep{shahrad2020serverless,fuerst2021faascache}.

A suspended request is not generating tokens, but retaining its key--value (KV) cache, the attention state computed for prior tokens, can still take resources away from active requests. Keeping the cache in GPU high-bandwidth memory (HBM) consumes serving capacity. Moving it to host dynamic random-access memory (DRAM) avoids that GPU holding cost but incurs transfer overhead, while discarding it eliminates the memory cost at the expense of recomputation when the request resumes. We summarize these tradeoffs as \textit{GPU opportunity cost}: the active-serving capacity forgone to preserve or restore suspended KV state.

At human-approval timescales, this tradeoff shifts substantially. Large-scale email data reports human-response latencies on the order of tens of minutes \citep{kooti2015email}, while human review of AI-agent-generated pull requests can extend to hours \citep{yu2026habituation}. We therefore use 30 minutes as a reference wait scale for asynchronous approval, rather than as a fixed wait assumption. At this timescale, the resume \textit{latency} becomes secondary (1s stall is 0.06\% of the wait), while the \textit{GPU opportunity cost} of retaining suspended state continues to accumulate throughout the wait.

It then follows that there is no single best host-retention timer for all approval workloads. If we were to model this regime using memoryless waits, such as an exponential distribution, the elapsed time is uninformative and expiration only adds churn. For heterogeneous short- and long-waiting populations, age can help separate contexts likely to resume soon from those likely to remain suspended.

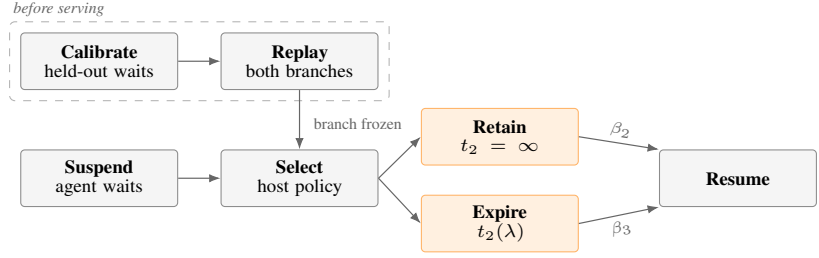
\begin{figure}[t]
\centering
\begin{tikzpicture}[
  font=\scriptsize,
  box/.style={draw=black!45, rounded corners=1.5pt, fill=black!4,
              align=center, inner sep=2.5pt, minimum height=7.5mm,
              text width=19mm},
  lit/.style={box, fill=orange!12, draw=orange!60},
  ar/.style={-{Latex[length=1.5mm]}, draw=black!60, line width=0.45pt},
  lbl/.style={font=\tiny, text=black!60, inner sep=1.5pt},
]
    \node[box] (cal) at (0,1.55)    {\textbf{Calibrate}\\[-1pt]held-out waits};
    \node[box] (rep) at (2.65,1.55) {\textbf{Replay}\\[-1pt]both branches};
    \draw[ar] (cal) -- (rep);
    \begin{scope}[on background layer]
      \node[draw=black!30, dashed, rounded corners=2pt, inner sep=4pt,
            fit=(cal)(rep)] (band) {};
    \end{scope}
    \node[lbl, anchor=south west] at (band.north west) {\textit{before serving}};
    
    \node[box] (susp) at (0,0)     {\textbf{Suspend}\\[-1pt]agent waits};
    \node[box] (sel)  at (2.65,0)  {\textbf{Select}\\[-1pt]host policy};
    \node[lit] (ret)  at (5.3,0.55){\textbf{Retain}\\[-1pt]$t_2=\infty$};
    \node[lit] (exp)  at (5.3,-0.6){\textbf{Expire}\\[-1pt]$t_2(\lambda)$};
    \node[box] (res)  at (8.45,0)  {\textbf{Resume}};
    \draw[ar] (susp) -- (sel);
    \draw[ar] (sel.east) -- (ret.west);
    \draw[ar] (sel.east) -- (exp.west);
    \draw[ar] (ret.east) -- node[lbl, above=0.5pt, sloped] {$\beta_2$} (res.north west);
    \draw[ar] (exp.east) -- node[lbl, below=0.5pt, sloped] {$\beta_3$} (res.south west);
    \draw[ar] (rep.south) -- (2.65,1.05) -- (sel.north);
    \node[lbl, anchor=west] at (2.78,0.72) {branch frozen};
\end{tikzpicture}
\caption{Regime-aware retention controller. Held-out waits select host-retain ($t_2=\infty$) or load-indexed expiration $t_2(\lambda)$ before serving; resume costs $\beta_2$ from host or $\beta_3$ after expiration (\S\ref{sec:policy}).}
\label{fig:controller}
\end{figure}

Thus, our approach selects between indefinite host retention and load-indexed expiration using a calibration sample, then fixes the choice before the held-out serving run. 

Our contributions are: \textbf{(i)} we quantify the cost of suspended-KV residency, showing a 41\% goodput loss from enforced residency and a roughly $10\times$ resume latency increase after eviction; \textbf{(ii)} we formulate a regime-aware retention policy using GPU opportunity cost to choose among HBM, host DRAM, and recomputation without per-request wait prediction; and \textbf{(iii)} under host-memory pressure, our controller improves active-request goodput by 23--51\% over the vLLM baselines, 22--29\% over MORI, and 41--52\% over Continuum.

\section{System-level characterization of human approval suspension}
\label{sec:regime}

We first characterize the two quantities that determine the retention decision at an approval gate: \textbf{(i)} how much reusable KV state remains when the agent pauses, and \textbf{(ii)} how much active-serving capacity is lost by keeping that state resident.

\paragraph{Approval-gated workload.} To measure the state left behind at an approval gate, we construct approval-gated requests from $\tau^2$-bench \citep{tau2bench}, which records multi-turn interactions between tool-using agents, simulated users, and task environments. Its retail and airline policies require explicit user confirmation before state-changing database actions. From the 2,624 released retail and airline runs, we retain trajectories containing a tool the environment marks \texttt{ToolType.WRITE} (e.g., order and reservation changes, cancellations, and returns) and truncate each immediately before the first such call, yielding 2,161 approval-gated prefixes.

The median suspended context contains 8.4K tokens: 5.4K shared system prefix and ${\sim}$3.0K private suffix. Prefix caching reuses the shared prefix \citep{vllm2026prefix}, so only the private suffix contributes per-request suspended state. Retaining this state is potentially valuable because 93--96\% of gates are approved by the benchmark's simulated user.\footnote{96.6\% over rule-classifiable requests; a manual audit of 50 unclassified requests yields a blended estimate of 93–96\%.}

\paragraph{The cost of KV residency.} The remaining question is what it costs the serving system to preserve this state during an approval wait. We call suspended KV \emph{resident} when its blocks remain allocated in HBM. In vLLM, suspended requests have no per-request residency guarantee: under memory pressure, their KV blocks may be \emph{evicted}, allowing that memory to be reclaimed \citep{kwon2023vllm}.

To characterize this behavior, we sweep $N=25$ to 72 suspended contexts, occupying 31--89\% of the KV pool. Under 3\,req/s of background load, which is the lowest rate at which eviction engages at all, residency falls from 100\% at $N=25$ to 10--14\% for $N\geq50$ (Table~\ref{tab:residency}). Because vLLM changes residency automatically under pressure, stock behavior cannot isolate the serving cost of residency.

We therefore add victim exclusion to vLLM's \texttt{BlockPool} and run a closed-loop experiment with 48 concurrent active clients, sufficient to saturate the node, while increasing the suspended set. Under saturation, capacity occupied by suspended KV appears directly as lost goodput. The patch is disabled for all unmodified-vLLM runs. Appendix~\ref{app:instrument} gives the residency measurement procedure, sweep selection, saturation test, and patch details.

\begin{figure}[h]
  \centering
  \includegraphics[width=0.75\columnwidth]{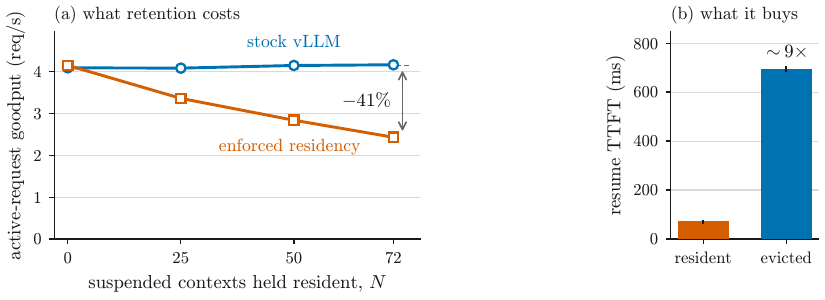}
  \caption{Suspended-KV residency tradeoff. \textbf{(a)} Enforced residency reduces goodput by 41\% at $N=72$. \textbf{(b)} Eviction increases resume TTFT from 66--75\,ms to 687--705\,ms ($\sim10\times$).}
  \label{fig:saturation}
\end{figure}

Figure~\ref{fig:saturation} shows the tradeoff. Unmodified vLLM keeps active-request \emph{goodput}, the rate of completed active requests, nearly flat by evicting most suspended KV, whereas enforced residency reduces goodput by 41\% at $N=72$. The cost instead appears at resume, where resident contexts have a time to first token (TTFT) of 66--75\,ms, compared with 687--705\,ms after eviction.

\section{Method}
\label{sec:pricing}
\label{sec:policy}

\subsection{Retention cost model}

Section~\ref{sec:regime} shows that retaining suspended KV trades active-serving capacity for faster resume. We place both costs on the same scale using \textit{GPU opportunity cost}, measured in GPU-seconds. A suspended context can be kept in HBM, moved to host DRAM, or discarded and recomputed when the request resumes.

For state $j \in \{1=\text{GPU HBM},\,2=\text{host DRAM},\, 3=\text{discard}\}$, let $\alpha_j$ denote its holding cost per second and $\beta_j$ its one-time resume cost. If a request remains in state $j$ for $\ell$ seconds, its cost is
\begin{equation}
  C_j(\ell)=\alpha_j\ell+\beta_j.
  \label{eq:cost-line}
\end{equation}

HBM incurs holding cost $\alpha_1$ while resident and has $\beta_1=0$. A host-resident copy pays restore cost $\beta_2$ at resume, while discarded state pays recomputation cost $\beta_3$.

\paragraph{Uncongested regime.}
When host DRAM has spare capacity, $\alpha_2=\alpha_3=0$: host DRAM consumes no direct GPU holding capacity, while discarded state occupies no serving memory.

\paragraph{Congested regime.}
Once host DRAM becomes scarce, retaining a host copy imposes an opportunity cost on other suspended contexts. We capture this with a nonzero, load-dependent holding price $\alpha_2(\lambda)$, where $\lambda$ is the offered rate of suspended requests. Section~\ref{sec:host-congestion} derives this price from $\lambda$ and the finite host capacity.

Thus, the cost model requires calibrating only three platform-specific coefficients: the HBM holding cost $\alpha_1$, host-resume cost $\beta_2$, and recomputation cost $\beta_3$. The remaining holding costs are either zero or derived from load.

For the HBM tier, two reference points determine when continued GPU residency is no longer worthwhile. In the uncongested regime, host offload becomes cheaper than HBM retention at
\begin{equation}
  t_1=\frac{\beta_2}{\alpha_1},
  \qquad
  t^{*}=\frac{\beta_3}{\alpha_1},
  \label{eq:hbm-break-even}
\end{equation}
where $t_1$ is the HBM--DRAM break-even and $t^{*}$ is the HBM--discard break-even. Host congestion is handled separately through the load-dependent DRAM--discard threshold $t_2(\lambda)$ in Section~\ref{sec:host-congestion}.

\begin{figure}[h]
  \centering
  \includegraphics[width=0.8\linewidth]{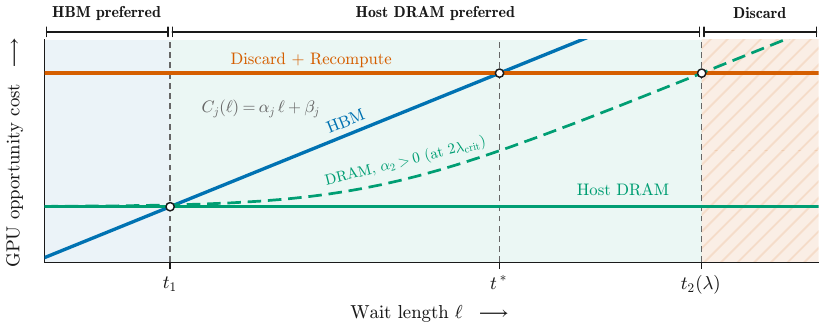}
  \caption{GPU opportunity cost across HBM, host DRAM, and discard; congestion introduces $\alpha_2(\lambda)$ and the DRAM--discard threshold $t_2(\lambda)$.}
  \label{fig:pricing-schematic}
\end{figure}

Figure~\ref{fig:pricing-schematic} summarizes the resulting cost structure: platform calibration determines the HBM holding and resume costs, while host congestion introduces the load-dependent DRAM holding price $\alpha_2(\lambda)$.

\subsection{Calibrating the price vector}
\label{sec:calibration}

The policy is calibrated to the serving platform rather than tied to fixed timing constants. We obtain the three opportunity-cost parameters from measurements of the state that is preserved or reconstructed at an approval gate: \textbf{(i) HBM holding cost $\alpha_1$}, derived from the fraction of the serving KV pool occupied by one suspended context and converted to node-level GPU-seconds; \textbf{(ii) host-resume cost $\beta_2$}, which captures GPU-side scheduling and synchronization overhead when restoring a host-resident copy, rather than DMA transfer time that can overlap GPU execution \citep{nvidia2026cuda}; and \textbf{(iii) recomputation cost $\beta_3$}, measured by re-prefilling the request-specific suffix with the shared prefix already cached.




These measurements produce the platform-specific price vector $(\alpha_1,\beta_2,\beta_3)$ used by the break-even rules above. Appendix~\ref{app:price-vector} gives full measurement protocol; Appendix~\ref{app:cross-platform} repeats the calibration on H100 NVL, A100 SXM, and L40S, showing how the calibrated parameters vary across hardware.

\subsection{Host-memory congestion}
\label{sec:host-congestion}

Host DRAM has negligible GPU holding cost while capacity is available. Once the tier is full, admitting a new suspended context requires discarding an existing host-resident copy, which later incurs the recomputation cost $\beta_3$.

Little's law relates the average number of resident requests $N$ to the arrival rate $\lambda$ and mean residence time $W$: $N=\lambda W$ \citep{littlelaw1961}. Here, it lets us translate a human-approval residence timescale into the offered load at which the finite host tier becomes saturated.

Let $C_2$ denote the number of suspended contexts that fit in host DRAM and $W_{\rm ref}$ the reference residence scale. By Little's law, host saturation occurs at $\lambda_{\mathrm{crit}}=C_2/W_{\rm ref}$. Above this threshold, we use a mean-field congestion price. Excess arrivals occur at rate $\lambda-\lambda_{\mathrm{crit}}$ under the reference normalization, and each overflow incurs recomputation cost $\beta_3$. Amortizing this cost across the $C_2$ occupied host slots gives
\begin{equation}
\begin{aligned}
\alpha_2(\lambda)
&=
\max(0,\lambda-\lambda_{\mathrm{crit}})
\frac{\beta_3}{C_2},
&\qquad
t_2(\lambda)
&=
\begin{cases}
\infty, & \lambda \le \lambda_{\mathrm{crit}},\\[1pt]
\dfrac{\beta_3-\beta_2}{\alpha_2(\lambda)},
& \lambda > \lambda_{\mathrm{crit}}.
\end{cases}
\end{aligned}
\label{eq:host-congestion}
\end{equation}

This yields a load-indexed host expiration policy, \texttt{cpu\_ttl}$(\lambda)$, which discards host-resident KV after $t_2(\lambda)$. Because the congestion price is a mean-field approximation rather than an exact occupancy model, we evaluate the resulting policy directly with finite-capacity replay in Appendix~\ref{app:load-sweep}.

\subsection{Regime-aware controller}
\label{sec:controller}

The cost model yields a simple two-stage retention policy. When a request suspends, its KV state moves to host DRAM; the HBM--DRAM break-even $t_1 \approx 1$s is short enough that we offload at suspension. The remaining decision is how long to keep the host copy.

A single host policy is not uniformly best across wait distributions. \emph{Host-retain} keeps the DRAM copy until the request resumes, whereas \texttt{cpu\_ttl}$(\lambda)$ discards it after the load-dependent threshold $t_2(\lambda)$. The latter can free scarce host capacity under heterogeneous long waits, but can also discard state that would have resumed soon.

We therefore select the host policy from an independent calibration sample before serving. We replay the same calibration waits under both modes, compute their total GPU opportunity cost, and freeze the lower-cost branch for the held-out run. At runtime the controller requires only the offered load $\lambda$ to evaluate $t_2(\lambda)$.

\section{Evaluation}
\label{sec:eval}

\subsection{Experimental setup}
\label{sec:eval:setup}

All live measurements use Llama-3.1-70B (BF16, TP4) on $4\times$H100 NVL GPUs with vLLM and a 375--380\,GB host-memory budget. We use live serving for end-to-end measurements and measured-cost replay for broader workload sweeps. Replay conditions contain 4,000 requests with Poisson arrivals at multiples of $\lambda_{\mathrm{crit}}$. We evaluate three human-scale wait families: a lognormal distribution ($\sigma = 1$), an exponential distribution, and a mixture of short approvals (exponential, mean 60 s, weight 0.5) and long waits (lognormal, $\sigma = 0.7$, mean 3,540 s). Each family is scaled to have mean wait $W_{\mathrm{ref}} = 1{,}800$ s. All end-to-end systems receive identical traces, and controller selection uses an independent calibration sample disjoint from the held-out serving trace.

Under our reference load normalization, host capacity is not binding below $\lambda_{\mathrm{crit}}$, so congestion-driven retention decisions are largely inactive. We therefore evaluate live serving at $2\lambda_{\mathrm{crit}}$, where the reference suspended load is twice the host-tier capacity ($\lambda W_{\rm ref}=2C_2$), making retention decisions consequential. Appendix~\ref{app:load-sweep} evaluates the controller against fixed host timers over a range of $0.5$--$3\times\lambda_{\mathrm{crit}}$.

\subsection{End-to-end serving performance}
\label{sec:eval:systems}

\begin{figure}[h]
  \centering
  \includegraphics[width=0.9\linewidth]{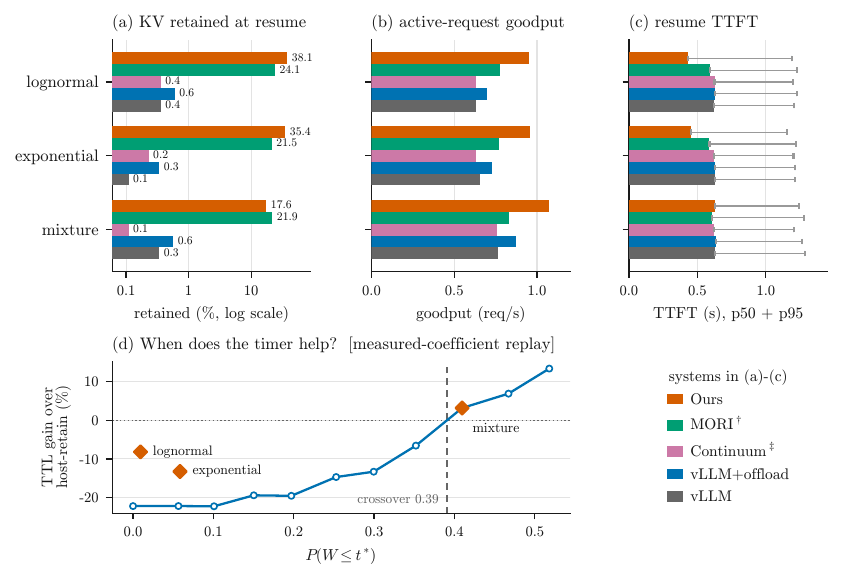}
  \caption{End-to-end serving at $2\lambda_{\mathrm{crit}}$: \textbf{(a)} KV retained at resume, \textbf{(b)} active-request goodput, and \textbf{(c)} resume TTFT (p50; p95 whiskers). \textbf{(d)} Measured-coefficient replay of \texttt{cpu\_ttl}$(\lambda)$ gain over host-retain; positive favors expiration, with crossover at $P(W\le t^{*})=0.39$. Appendix~\ref{app:eval-details} gives exact values. $^\dagger$MORI policy port; $^\ddagger$Continuum implementation details in Appendix~\ref{app:serving-config}.}
  \label{fig:e2e}
\end{figure}

We first ask whether retention policy improves active-serving goodput. We compare our regime-aware controller with unmodified vLLM \citep{kwon2023vllm}, vLLM with host offload enabled, MORI \citep{mori2026}, and Continuum \citep{continuum2025}.\footnote{MORI has no released implementation, so we evaluate its published policy through a port on vLLM. For Continuum, we use the first author's public fork \citep{continuum_code}, whose README states that it omits the paper's pause-duration estimator.} Figure~\ref{fig:e2e} reports live measurements at $2\lambda_{\mathrm{crit}}$.

\paragraph{Retention policy matters.} Unmodified vLLM retains only 0.4\% and 0.3\% of suspended KV at resume on the lognormal and mixture workloads, respectively. Enabling 380\,GB of host offload raises retention to only 0.6\% on both workloads, whereas our controller retains 38.1\% and 17.6\%. The difference is policy rather than capacity: under recency-based eviction \citep{kwon2023vllm}, contexts suspended for long human waits become natural eviction candidates. Our controller instead preserves suspended state selectively, yielding 23--51\% higher active-request goodput than the two vLLM configurations and 41--52\% higher goodput than Continuum.

\paragraph{More retention does not necessarily improve serving.} On the mixture workload, MORI retains more suspended KV than our controller (21.9\% versus 17.6\%) but achieves only 0.83\,req/s versus 1.07\,req/s, giving our controller 29\% higher goodput. MORI places its host-tier boundary at available capacity \citep{mori2026}, whereas our controller can expire a host copy at $t_2(\lambda)$ once its retention cost exceeds the cost of recomputation. The result is the intended behavior of the pricing model: retain enough state to avoid unnecessary recomputation without sacrificing active-serving capacity merely to maximize KV retention.

\subsection{Controller selection}

The end-to-end results show that neither retaining more KV nor expiring it aggressively is uniformly best. We therefore ask when \texttt{cpu\_ttl}$(\lambda)$ is preferable to host-retain. Using $P(W\le t^{*})$ to summarize the fraction of waits below the platform-calibrated break-even timescale, Figure~\ref{fig:e2e}(d) shows the gain of \texttt{cpu\_ttl}$(\lambda)$ over host-retain at fixed load. Expiration hurts when few requests resume before $t^{*}$, but becomes beneficial as the workload contains more short waits. The crossover occurs at $P(W\le t^{*})=0.39$ in this setup.

\section{Discussion}
\label{sec:discussion}

\paragraph{Related work.} Prior paused-agent systems use request-level signals such as reload cost, predicted pause duration, idleness, or workflow position \citep{infercept2024,continuum2025,mori2026,kvflow2025}. We instead target minute-to-hour approval waits and price retention by system load and finite host capacity; Appendix~\ref{app:related} gives a detailed comparison.

\paragraph{Limitations.} Our experiments use the configured offered load $\lambda$, so a production deployment would require online load estimation and periodic recalibration under changing approval-wait distributions. Approval outcomes are LLM-simulated and wait durations are synthetic, and we do not yet have a production trace showing how often deployments cross $\lambda_{\mathrm{crit}}$. MORI is evaluated through a policy port rather than its native serving stack. Our Continuum comparison uses the first author's public scheduling fork, which omits the paper's pause-duration estimator, and therefore does not constitute a full-system evaluation of Continuum. Our model-level measurements use a single model family. These results therefore establish a measurement-backed retention policy, rather than a complete production controller.

Overall, human-approval suspension turns KV retention into a capacity-allocation problem: residency consumes active-serving capacity, while eviction slows resume. Our results show that platform-calibrated, load-aware retention can navigate this tradeoff without per-request wait prediction, outperforming vLLM and prior agent-serving baselines in active-request goodput.

\clearpage

\bibliographystyle{plainnat}
\bibliography{refs}

@misc{claudecode2025,
  title        = {Claude Code},
  author       = {{Anthropic}},
  year         = {2025},
  howpublished = {\url{https://claude.com/product/claude-code}},
  note         = {Product page; accessed 2026-08-23}
}

@misc{langgraph2026,
  title        = {Interrupts},
  author       = {{LangChain}},
  year         = {2026},
  howpublished = {\url{https://docs.langchain.com/oss/python/langgraph/interrupts}},
  note         = {LangGraph documentation; accessed 2026-08-23}
}

@misc{kvflow2025,
  title         = {{KVFlow}: Efficient Prefix Caching for Accelerating {LLM}-Based Multi-Agent Workflows},
  author        = {Pan, Zaifeng and Patel, Ajjkumar and Hu, Zhengding and Shen, Yipeng and Guan, Yue and Li, Wan-Lu and Qin, Lianhui and Wang, Yida and Ding, Yufei},
  year          = {2025},
  eprint        = {2507.07400},
  archivePrefix = {arXiv},
  note          = {arXiv:2507.07400}
}

@misc{mori2026,
  title         = {Idleness is Relative: Exploiting Tool-Call Idle Windows for Offloading in Agentic Systems with {MORI}},
  author        = {Xia, Tian and Li, Hanchen and Li, Zhifei and Chen, Xiaokun and Kang, Hao and Qiao, Yifan and Xu, Yi and Stoica, Ion},
  year          = {2026},
  eprint        = {2606.00866},
  archivePrefix = {arXiv},
  note          = {arXiv:2606.00866}
}

@misc{continuum2025,
  title         = {Continuum: Efficient and Robust Multi-Turn {LLM} Agent Scheduling with {KV} Cache Time-to-Live},
  author        = {Li, Hanchen and He, Runyuan and Mang, Qiuyang and Zhang, Qizheng and Mao, Huanzhi and Chen, Xiaokun and Zhou, Hangrui and Cheung, Alvin and Gonzalez, Joseph and Stoica, Ion},
  year          = {2025},
  eprint        = {2511.02230},
  archivePrefix = {arXiv},
  note          = {arXiv:2511.02230}
}

@inproceedings{infercept2024,
  title     = {{InferCept}: Efficient Intercept Support for Augmented Large Language Model Inference},
  author    = {Abhyankar, Reyna and He, Zijian and Srivatsa, Vikranth and Zhang, Hao and Zhang, Yiying},
  booktitle = {Proceedings of the 41st International Conference on Machine Learning ({ICML})},
  year      = {2024},
  note      = {arXiv:2402.01869}
}

@misc{autellix2025,
  title         = {Autellix: An Efficient Serving Engine for {LLM} Agents as General Programs},
  author        = {Luo, Michael and Shi, Xiaoxiang and Cai, Colin and Zhang, Tianjun and Wong, Justin and Wang, Yichuan and Wang, Chi and Huang, Yanping and Chen, Zhifeng and Gonzalez, Joseph E. and Stoica, Ion},
  year          = {2025},
  eprint        = {2502.13965},
  archivePrefix = {arXiv},
  note          = {arXiv:2502.13965}
}

@inproceedings{kwon2023vllm,
  title     = {Efficient Memory Management for Large Language Model Serving with {PagedAttention}},
  author    = {Kwon, Woosuk and Li, Zhuohan and Zhuang, Siyuan and Sheng, Ying and Zheng, Lianmin and Yu, Cody Hao and Gonzalez, Joseph E. and Zhang, Hao and Stoica, Ion},
  booktitle = {Proceedings of the 29th {ACM} Symposium on Operating Systems Principles ({SOSP})},
  pages     = {611--626},
  year      = {2023}
}

@misc{tau2bench,
  title         = {{$\tau^2$}-Bench: Evaluating Conversational Agents in a Dual-Control Environment},
  author        = {Barres, Victor and Dong, Honghua and Ray, Soham and Si, Xujie and Narasimhan, Karthik},
  year          = {2025},
  eprint        = {2506.07982},
  archivePrefix = {arXiv},
  note          = {arXiv:2506.07982}
}

@inproceedings{shahrad2020serverless,
  title     = {Serverless in the Wild: Characterizing and Optimizing the Serverless Workload at a Large Cloud Provider},
  author    = {Shahrad, Mohammad and Fonseca, Rodrigo and Goiri, {\'I}{\~n}igo and Chaudhry, Gohar and Batum, Paul and Cooke, Jason and Laureano, Eduardo and Tresness, Colby and Russinovich, Mark and Bianchini, Ricardo},
  booktitle = {2020 {USENIX} Annual Technical Conference ({USENIX} {ATC} 20)},
  pages     = {205--218},
  year      = {2020}
}

@inproceedings{fuerst2021faascache,
  title     = {{FaasCache}: Keeping Serverless Computing Alive with Greedy-Dual Caching},
  author    = {Fuerst, Alexander and Sharma, Prateek},
  booktitle = {Proceedings of the 26th {ACM} International Conference on Architectural Support for Programming Languages and Operating Systems ({ASPLOS})},
  pages     = {386--400},
  year      = {2021}
}

@article{littlelaw1961,
  title   = {A Proof for the Queuing Formula: {$L = \lambda W$}},
  author  = {Little, John D. C.},
  journal = {Operations Research},
  volume  = {9},
  number  = {3},
  pages   = {383--387},
  year    = {1961}
}

@inproceedings{kooti2015email,
  title     = {Evolution of Conversations in the Age of Email Overload},
  author    = {Kooti, Farshad and Aiello, Luca Maria and Grbovic, Mihajlo
               and Lerman, Kristina and Mantrach, Amin},
  booktitle = {Proceedings of the 24th International Conference on World Wide Web},
  year      = {2015}
}

@article{yu2026habituation,
  title   = {Habituation at the Gate: Rising Approval and Declining Scrutiny
             in Human Review of AI Agent Code},
  author  = {Yu, Haoran and Liu, Lifei and Jiang, Xiaochong and Jia, Yuwen
             and Wang, Su and Qian, Pin and Chen, Yihang},
  journal = {arXiv preprint arXiv:2606.22721},
  year    = {2026}
}

@misc{githubcopilot2026,
  author       = {{GitHub}},
  title        = {Configuring GitHub Copilot CLI},
  year         = {2026},
  howpublished = {\url{https://docs.github.com/en/copilot/how-tos/copilot-cli/set-up-copilot-cli/configure-copilot-cli}},
  note         = {Accessed 2026-08-27}
}

@inproceedings{wu2024autogen,
  title     = {AutoGen: Enabling Next-Gen LLM Applications via Multi-Agent Conversation},
  author    = {Wu, Qingyun and Bansal, Gagan and Zhang, Jieyu and Wu, Yiran
               and Li, Beibin and Zhu, Erkang and Jiang, Li and Zhang, Xiaoyun
               and Zhang, Shaokun and Liu, Jiale and Awadallah, Ahmed Hassan
               and White, Ryen W. and Burger, Doug and Wang, Chi},
  booktitle = {Conference on Language Modeling (COLM)},
  year      = {2024}
}

@misc{openaicodexapp2026,
  author       = {{OpenAI}},
  title        = {Introducing the Codex App},
  year         = {2026},
  howpublished = {\url{https://openai.com/index/introducing-the-codex-app/}},
  note         = {Accessed 2026-08-27}
}

@manual{nvidia2026cuda,
  title        = {{CUDA Programming Guide}},
  author       = {{NVIDIA Corporation}},
  year         = {2026},
  url          = {https://docs.nvidia.com/cuda/cuda-programming-guide/},
  note         = {Accessed: 2026-08-29}
}

@misc{vllm2026prefix,
  title        = {{Automatic Prefix Caching}},
  author       = {{vLLM Project}},
  year         = {2026},
  howpublished = {\url{https://docs.vllm.ai/en/latest/features/automatic_prefix_caching/}},
  note         = {Accessed: 2026-08-29}
}

@misc{continuum_code,
  author       = {Hanchen Li},
  title        = {vLLM with Continuum Scheduling},
  year         = {2026},
  howpublished = {\url{https://github.com/Hanchenli/vllm-continuum}},
  note         = {Commit 316a587, accessed August 2026}
}

\appendix

\section{Residency-enforcement instrument}
\label{app:instrument}

\begin{table}[h]
  \caption{Realized HBM residency of $N$ suspended Llama-3.1-70B contexts
  under 3\,req/s background load (TP4, $4\times$H100 NVL).}
  \label{tab:residency}
  \centering
  \small
  \begin{tabular}{@{}lcccc@{}}
    \toprule
    Residency (\%) & $N{=}25$ & $N{=}50$ & $N{=}65$ & $N{=}72$ \\
    \midrule
    unmodified vLLM & 100 & \multicolumn{3}{c}{10--14} \\
    enforced residency & 100 & 100 & 98 & 93 \\
    \bottomrule
  \end{tabular}
\end{table}

\paragraph{Residency measurement and sweep range.} Each entry in Table~\ref{tab:residency} is the fraction of suspended contexts whose KV state remains in HBM when the request resumes. We determine residency from resume TTFT, using per-context resident and evicted reference measurements, rather than from internal engine counters.

We choose $N=\{25,50,65,72\}$ to span the usable KV-capacity range. A full gate context occupies 2.75\,GB, so these points correspond to approximately 31\%, 62\%, 80\%, and 89\% of the measured 680,768-token KV pool. $N=25$ provides an uncontended anchor where unmodified vLLM retains all contexts, while $N=72$ is near the largest suspended set that still leaves capacity for active requests. Around $N\approx81$, suspended state alone fills the pool. We use 3\,req/s of background load because it is the lowest tested rate at which eviction becomes visible; at 1.5\,req/s, both arms retain 97--100\% suspended contexts across the sweep.

Under pressure, unmodified vLLM residency drops to 10--14\%, whereas victim exclusion maintains 93--100\% residency. The small shortfall at $N=72$ reflects admission throttling as free KV capacity approaches zero, rather than eviction of pinned contexts.

\paragraph{Victim-exclusion patch.} To vary residency independently of memory pressure, we modify vLLM's \texttt{BlockPool} so that selected suspended contexts cannot be chosen as eviction victims. Pinning removes their blocks from the doubly linked LRU free queue in $O(1)$ time, and unpinning returns them to the eviction tail. Pinned blocks therefore remain unavailable to active requests, allowing their capacity cost to appear naturally through the serving scheduler. The patch is disabled for all unmodified-vLLM runs, and a regression run confirms equivalence with the stock engine when victim exclusion is disabled.

\paragraph{Choosing the active load.} The closed-loop experiment must operate near saturation; otherwise, KV held by suspended requests can consume unused headroom without reducing measured goodput. We therefore sweep the number of concurrent active clients $C$ at $N=0$ before fixing the load used in Figure~\ref{fig:saturation}. Each point uses a 90\,s window, 64-token completions, and the same active-request prompt pool. Table~\ref{tab:concurrency-sweep} reports this sweep.

\begin{table}[h]
  \caption{Concurrency sweep used to select the closed-loop operating point.}
  \label{tab:concurrency-sweep}
  \centering
  \small
  \begin{tabular}{@{}rrrr@{}}
    \toprule
    $C$ & goodput (req/s) & TTFT p50 (s) & TTFT p95 (s) \\
    \midrule
     8 & 1.70 & 0.68 &  1.8 \\
    16 & 2.01 & 0.85 &  2.5 \\
    32 & 2.27 & 1.11 & 14.8 \\
    \textbf{48} & \textbf{2.42} & 1.67 & 28.1 \\
    64 & 2.32 & 2.31 & 39.4 \\
    \bottomrule
  \end{tabular}
\end{table}

Goodput peaks at $C=48$ and declines at $C=64$ while tail latency continues to increase, indicating that the node has entered the saturated regime. We therefore fix $C=48$ for the residency-cost sweep in Figure~\ref{fig:saturation}. At this operating point, KV capacity withheld by suspended contexts competes directly with active requests and can therefore be measured as lost goodput.

\section{Price-vector calibration}
\label{app:price-vector}

We calibrate the price vector on Llama-3.1-70B (bf16, TP4) running on $4\times$H100 NVL GPUs, using an exclusive server with no background load. Per-token KV occupies $80\times8\times128\times2\times2=327{,}680$ bytes, so the median 3.0K-token private suffix occupies 0.98\,GB, compared with 2.75\,GB for the full 8.4K-token gate context.

\paragraph{$\alpha_1$: HBM holding cost.} $\alpha_1$ is a capacity-accounting price rather than a timing measurement. The median suffix occupies $3{,}000/680{,}768$ of the measured TP4 KV pool. Charging this fraction across four GPU ranks gives
\[
  \alpha_1
  =4\times\frac{3{,}000}{680{,}768}
  =0.0176\ \text{GPU-s/s}.
\]
The end-to-end engine reports a 667,632-token KV pool, 1.9\% below the calibration run. Independently, the closed-loop experiment in Section~\ref{sec:regime} gives an empirical slope of $0.023$--$0.030$ GPU-s/s, bracketing this accounting price from above.

\paragraph{$\beta_2$: host-resume cost.} We benchmark the larger 2.75\,GB full-context payload to upper-bound transfer interference, although the state attributable to one suspended request is the 0.98\,GB private suffix. Copies use page-locked host memory and \texttt{torch.Tensor.copy\_(non\_blocking=True)}, with synchronization before and after each measurement. Across 10 repetitions, median device-to-host and host-to-device times are 0.0480\,s (57.3\,GB/s) and 0.0478\,s (57.5\,GB/s), respectively.

Because these transfers are DMA-driven, transfer wall time is not charged one-for-one as GPU opportunity cost. We assign $\beta_2=0.02$ GPU-s as a conservative allowance for GPU-side scheduling and synchronization overhead. Setting this allowance to zero changes $\beta_3-\beta_2$ by about 1\%; charging the full 0.0478\,s H2D interval across all four GPUs ($0.19$ GPU-s) changes it by under 10\%.

\paragraph{$\beta_3$: recomputation cost.} We measure suffix re-prefill through the serving API. For 30 gate prompts sampled with a fixed seed (lengths $\leq32$K), we reset the prefix cache, warm the shared system prefix, and then issue the full gate prompt with no competing traffic. Each prompt is repeated three times (90 measurements). With the shared prefix cached, the median suffix-resume TTFT is 0.479\,s and a cold full-context run takes 1.366\,s.

Because TTFT is measured client-side, it also includes API and tokenization overhead. We therefore use it as a conservative end-to-end charge rather than a pure GPU timing measurement. Charging the observed 0.479\,s interval across four TP ranks gives
\[
  \beta_3 = 0.479\times4 = 1.91\ \text{GPU-s}.
\]
\paragraph{Full-context sensitivity.} The main HBM price charges only the private suffix because the shared prefix is reused. Charging the full 8.4K-token context instead gives $\alpha_1^{\mathrm{full}}=0.0494$ GPU-s/s; we use this only as a sensitivity bound.

\section{Cross-platform ablation}
\label{app:cross-platform}

To test whether the calibration recipe transfers across hardware, we repeat the measurements from Appendix~\ref{app:price-vector} on H100 NVL, A100 SXM, and L40S platforms, then replay the controller using each platform's own calibrated price vector. All experiments use Llama-3.1-70B in bf16 with TP4.

For each platform, we measure host-to-device bandwidth, suffix re-prefill cost $\beta_3$, and KV-pool capacity, then derive $\alpha_1$, $t_1$, and $t^*$ using the same procedure as in Appendix~\ref{app:price-vector}. We replay host-retain and \texttt{cpu\_ttl}$(\lambda)$ at $2\lambda_{\mathrm{crit}}$ using the same wait families and a fixed host-tier capacity of $C_2=425$.

\begin{table}[h]
\caption{Cross-platform calibration and replay at $2\lambda_{\mathrm{crit}}$. Each platform uses its own measured price vector and derived break-even times. The replay columns report the relative cost of \texttt{cpu\_ttl}$(\lambda)$ versus host-retain; negative values favor expiration.}
\label{tab:platforms}
\centering
\footnotesize
\setlength{\tabcolsep}{3.2pt}
\begin{tabular}{@{}l rrrr rrr rrrr@{}}
\toprule
 & \multicolumn{4}{c}{measured}
 & \multicolumn{3}{c}{derived}
 & \multicolumn{4}{c}{replay: TTL vs.\ retain} \\
\cmidrule(lr){2-5}
\cmidrule(lr){6-8}
\cmidrule(l){9-12}
Platform
& \makecell[r]{PCIe\\GB/s}
& \makecell[r]{$\beta_3$\\GPU-s}
& \makecell[r]{KV pool\\tokens}
& \makecell[r]{HBM\\GB}
& \makecell[r]{$\alpha_1$\\GPU-s/s}
& \makecell[r]{$t_1$\\s}
& \makecell[r]{$t^{*}$\\s}
& lognormal
& exponential
& mixture
& \makecell[r]{crossover\\$P(W\le t^{*})$} \\
\midrule
H100 NVL (Gen5)
& 57.5 & 1.91 & 680{,}768 & 4$\times$94
& 0.0176 & 1.13 & 109
& $+8.7\%$ & $+12.9\%$ & $-3.1\%$ & 0.39 \\

A100 SXM (Gen4)
& 26.2 & 2.61 & 516{,}976 & 4$\times$80
& 0.0232 & 0.86 & 112
& $+8.8\%$ & $+14.0\%$ & $-3.1\%$ & 0.39 \\

L40S (Gen4)$^{*}$
& 27.0 & 5.14 & 101{,}504 & 4$\times$48
& 0.118 & 0.17 & 43.5
& $+9.1\%$ & $+15.0\%$ & $-3.1\%$ & 0.23 \\
\bottomrule
\end{tabular}

\par\smallskip
{\scriptsize
$^{*}$On L40S, compute becomes limiting before KV capacity over part of the measured range; the reported $\alpha_1$ therefore corresponds to the KV-bound regime. We hold $C_2=425$ fixed across platforms to isolate GPU platform effects. \par}
\end{table}

The calibrated constants vary substantially across platforms: $\alpha_1$ changes by nearly $7\times$, while $t^*$ ranges from 43.5\,s to 112\,s. Despite these shifts, the policy ordering is unchanged: host-retain remains preferred for the lognormal and exponential waits, while \texttt{cpu\_ttl}$(\lambda)$ remains preferred for the mixture. The result suggests that the calibration procedure transfers across platforms, while the resulting thresholds themselves should be re-measured rather than reused across hardware.

\section{Exact semantics of \texttt{cpu\_ttl}$(\lambda)$}
\label{app:semantics}

The main text uses $W_{\rm ref}=1{,}800$\,s to define $\lambda_{\mathrm{crit}}=C_2/W_{\rm ref}$, and every evaluated wait family is scaled to mean $W_{\rm ref}$ so that "$2\lambda_{\mathrm{crit}}$" denotes the same load for all three. For a wait distribution $W$, host-retain has mean DRAM occupancy $\lambda\mathbb{E}[W]$, while a TTL policy has occupancy $\lambda\mathbb{E}[\min(W,t_2)]$. Replay enforces the finite $C_2$ host slots directly using the sampled waits.

The TTL decision is one-shot. On suspension, a context moves to host DRAM and receives $t_2(\lambda)$. If the request has not resumed when the timer fires, the host copy is discarded and the request is not subsequently re-scored from elapsed wait. In our experiments, $\lambda$ is the configured offered rate for the condition.

Host capacity is enforced at admission. In replay, a suspended context that arrives when a host slot is available is admitted; one that finds the tier full is not admitted and recomputes at resume, and residents are never displaced. In the live controller the tier admits every new context and evicts by priority: blocks whose timer has expired, then transient blocks of active requests, then least-recently-used suspended blocks. Both rules are identical under host-retain and \texttt{cpu\_ttl}$(\lambda)$.

The two branches differ only in proactive expiration: \texttt{cpu\_ttl}$(\lambda)$ may discard a resident context when its $t_2(\lambda)$ timer fires, while host-retain does not. TTL expiration can therefore free host capacity before a later arrival would otherwise trigger replacement.

\section{Extended related work}
\label{app:related}

\begin{table}[h]
  \caption{Retention signals in prior systems. Prior agent-serving policies primarily use request-level signals, whereas ours indexes retention on system load. $^{\dagger}$Autellix names human input as an interrupt type but scopes it out of serving.}
  \label{tab:signals}
  \centering
  \small
  \begin{tabular}{llll}
    \toprule
    System & Idle source & Design scale & Retention signal \\
    \midrule
    vLLM \citep{kwon2023vllm}
      & memory pressure & -- & recency (LRU) \\
    InferCept \citep{infercept2024}
      & tool call & seconds & reload cost \\
    Autellix \citep{autellix2025}
      & tool call, human$^{\dagger}$ & seconds & none \\
    KVFlow \citep{kvflow2025}
      & tool call & seconds & workflow position \\
    MORI \citep{mori2026}
      & tool call & seconds & idleness rank \\
    Continuum \citep{continuum2025}
      & tool call & $\leq2$\,s & predicted duration \\
    \texttt{cpu\_ttl}$(\lambda)$ (ours)
      & approval gate & minutes--hours & system load $\lambda$ \\
    \bottomrule
  \end{tabular}
\end{table}

\paragraph{MORI.} MORI uses the same broad placement choices (GPU, host memory, and recomputation) but ranks requests using relative idleness over a short history window \citep{mori2026}. In our 2,161 approval-gated trajectories, the p5--p95 spread of this statistic contracts from 0.359 at tool-call-scale waits to 0.018 at half-hour waits. This motivates using system pressure rather than request-level idleness as the control signal at human-approval timescales.

\paragraph{Continuum.} Continuum indexes retention on a predicted pause-duration bound \citep{continuum2025}. Its published system primarily targets seconds-scale tool calls. We evaluate its implementation under the same human-scale wait workloads used in our end-to-end experiments.

\paragraph{InferCept.} InferCept chooses among preserving, swapping, and recomputing intercepted KV using expected GPU-memory waste \citep{infercept2024}. Its measurements target seconds-scale interceptions. At longer waits, continuous GPU retention becomes increasingly expensive, leaving the additional question studied here: when a finite host tier should itself release suspended state under load.

\paragraph{Serverless keep-alive.} Serverless systems face a related retain-or-release decision, using per-function idle-time distributions or reuse value to choose container keep-alive behavior \citep{shahrad2020serverless,fuerst2021faascache}. Our setting instead prices retention against shared accelerator and host-memory capacity.

\section{Extended evaluation}
\label{app:eval-details}

\subsection{Serving configuration}
\label{app:serving-config}

All requests use greedy decoding (\texttt{temperature=0}). A gate request generates one token at suspension and eight tokens at resume. Each of the $C=48$ closed-loop active clients draws a prompt uniformly from a 1,160-prompt background pool and generates 64 tokens per request. Gate contexts are drawn from a disjoint 1,000-prompt pool.

For the vLLM-based systems, we use vLLM v0.27.1 with \nolinkurl{--tensor-parallel-size} 4, \nolinkurl{--max-model-len} 32768, and \nolinkurl{--gpu-memory-utilization} 0.92. The host tier uses \nolinkurl{OffloadingConnector} with \nolinkurl{kv_role=kv_both}, \nolinkurl{offload_prompt_only=false}, and \nolinkurl{cpu_bytes_to_use}$=380\times10^9$ bytes ($375\times10^9$ for the exponential workload). These settings are identical across unmodified vLLM, vLLM+offload, the MORI policy port, and our controller; only \nolinkurl{eviction_policy} and \nolinkurl{cache_policy_module_path} differ where applicable.

Continuum uses the public fork released by the first author (vLLM 0.10.2 base) \citep{continuum_code}, which implements Continuum scheduling without the paper's pause-duration estimator, so we use its shipped 2\,s pin TTL.

\subsection{Exact end-to-end measurements}

Table~\ref{tab:e2e-exact} reports the values underlying Figure~\ref{fig:e2e}.

\begin{table}[!h]
  \centering
  \small
  \caption{Exact live end-to-end measurements underlying
  Figure~\ref{fig:e2e}. Retained, recomputed, and resume TTFT are measured
  after draining the serving window.}
  \label{tab:e2e-exact}
  \begin{tabular}{llrrrr}
    \toprule
    System & Waits &
    Goodput &
    Retained &
    Recomputed &
    Resume TTFT \\
    & & (req/s) & (\%) & (\%) & (ms) \\
    \midrule
    Ours & lognormal   & 0.95 & 38.1 & 61.4 & 435 \\
         & exponential & 0.96 & 35.4 & 63.8 & 455 \\
         & mixture     & 1.07 & 17.6 & 81.8 & 628 \\
    \addlinespace[2pt]
    MORI$^\dagger$ & lognormal & 0.78 & 24.1 & 75.9 & 590 \\
         & exponential & 0.77 & 21.5 & 78.5 & 588 \\
         & mixture     & 0.83 & 21.9 & 77.5 & 606 \\
    \addlinespace[2pt]
    Continuum$^\ddagger$ & lognormal & 0.63 & 0.4 & 98.6 & 625 \\
         & exponential & 0.63 & 0.2 & 98.8 & 620 \\
         & mixture     & 0.76 & 0.1 & 98.8 & 620 \\
    \addlinespace[2pt]
    vLLM+offload & lognormal & 0.70 & 0.6 & 95.4 & 626 \\
         & exponential & 0.73 & 0.3 & 96.1 & 630 \\
         & mixture     & 0.87 & 0.6 & 95.5 & 633 \\
    \addlinespace[2pt]
    vLLM & lognormal   & 0.63 & 0.4 & 94.4 & 619 \\
         & exponential & 0.66 & 0.1 & 96.0 & 630 \\
         & mixture     & 0.77 & 0.3 & 96.1 & 628 \\
    \bottomrule
  \end{tabular}

  \par\smallskip
  {\scriptsize $^\dagger$Policy port on the same serving engine. $^\ddagger$First author's public fork (vLLM 0.10.2 base) as shipped, fixed 2\,s pin TTL; scored against its own resident/evicted references.}
\end{table}

For each resumed context, we normalize its TTFT between context-specific resident and evicted references:
\[
f =
\frac{\mathrm{TTFT}-\mathrm{TTFT}_{\mathrm{hit}}}
{\mathrm{TTFT}_{\mathrm{cold}}-\mathrm{TTFT}_{\mathrm{hit}}},
\]
clipped to $[0,1]$. \emph{Retained} is the fraction of contexts with $f\le0.15$, while \emph{Recomputed} is the mean $f$, used as a normalized estimate of recomputation at resume. The two columns measure different statistics and are not expected to sum to 100\%.

\FloatBarrier

\subsection{Load-sweep replay}
\label{app:load-sweep}

To test whether the controller requires per-load tuning, we replay 12
conditions spanning $0.5$--$3\times\lambda_{\mathrm{crit}}$ at $C_2=425$.
We compare \texttt{cpu\_ttl}$(\lambda)$ against fixed host-retention policies
that offload at suspension and discard after 10, 30, or 60 minutes, or never
($\infty$, host-retain). The controller is calibrated once per wait family,
after which the selected branch is fixed across all loads.

Table~\ref{tab:load-sweep} reports the resulting GPU opportunity cost.
The controller matches the best fixed policy throughout the lognormal and
exponential workloads. On the mixture workload, it remains within 0.2\% of
the best fixed policy at $2\lambda_{\mathrm{crit}}$ and outperforms it by
3.6\% at $3\lambda_{\mathrm{crit}}$. Thus, the calibration-selected branch
remains competitive across load without per-load TTL tuning.

\begin{table}[h]
  \centering
  \small
  \caption{Load-sweep replay at $C_2=425$. Entries are GPU opportunity cost per request (GPU-s; lower is better). The controller entry shows the calibration-selected branch in parentheses. Gain is relative to the best fixed policy for each condition; positive favors the controller. Bold marks the lowest cost in each row.}
  \label{tab:load-sweep}

  \begin{tabular}{llrrrrrrr}
    \toprule
    Waits &
    $\lambda/\lambda_{\mathrm{crit}}$ &
    \texttt{cpu\_ttl}$(\lambda)$ &
    $\infty$ &
    10 min &
    30 min &
    60 min &
    Controller &
    Gain \\
    \midrule

    lognormal
      & 0.5
      & \textbf{0.020}
      & \textbf{0.020}
      & 1.374
      & 0.598
      & 0.243
      & \textbf{0.020 (retain)}
      & 0.0\% \\
      & 1
      & \textbf{0.065}
      & \textbf{0.065}
      & 1.374
      & 0.598
      & 0.243
      & \textbf{0.065 (retain)}
      & 0.0\% \\
      & 2
      & 0.851
      & \textbf{0.787}
      & 1.374
      & 0.868
      & 0.834
      & \textbf{0.787 (retain)}
      & 0.0\% \\
      & 3
      & 1.239
      & \textbf{1.081}
      & 1.375
      & 1.199
      & 1.138
      & \textbf{1.081 (retain)}
      & 0.0\% \\

    \addlinespace[2pt]

    exponential
      & 0.5
      & \textbf{0.020}
      & \textbf{0.020}
      & 1.373
      & 0.737
      & 0.278
      & \textbf{0.020 (retain)}
      & 0.0\% \\
      & 1
      & \textbf{0.132}
      & \textbf{0.132}
      & 1.373
      & 0.737
      & 0.278
      & \textbf{0.132 (retain)}
      & 0.0\% \\
      & 2
      & 0.993
      & \textbf{0.877}
      & 1.373
      & 0.999
      & 0.937
      & \textbf{0.877 (retain)}
      & 0.0\% \\
      & 3
      & 1.307
      & \textbf{1.145}
      & 1.373
      & 1.266
      & 1.229
      & \textbf{1.145 (retain)}
      & 0.0\% \\

    \addlinespace[2pt]

    mixture
      & 0.5
      & \textbf{0.020}
      & \textbf{0.020}
      & 0.957
      & 0.728
      & 0.363
      & \textbf{0.020 (ttl)}
      & 0.0\% \\
      & 1
      & \textbf{0.072}
      & \textbf{0.072}
      & 0.957
      & 0.728
      & 0.363
      & \textbf{0.072 (ttl)}
      & 0.0\% \\
      & 2
      & 0.759
      & 0.785
      & 0.957
      & \textbf{0.758}
      & 0.789
      & 0.759 (ttl)
      & $-0.2$\% \\
      & 3
      & \textbf{0.923}
      & 1.073
      & 0.957
      & 1.082
      & 1.126
      & \textbf{0.923 (ttl)}
      & $+3.6$\% \\

    \bottomrule
  \end{tabular}
\end{table}

\section{Compute resources}
\label{app:compute}

\paragraph{Hardware.}
Live experiments use a single node with $4\times$NVIDIA H100 NVL GPUs
(94\,GB HBM each), 128 vCPUs, and 1.5\,TiB host DRAM. The available
shared-memory budget limits the host tier to 375--380\,GB. The closed-loop
residency experiment uses identically configured
nodes. Replay, trace generation, and plotting are CPU-only and complete
within minutes.

\begin{table}[h]
\centering
\small
\caption{Approximate compute for the experiments reported in the paper.}
\label{tab:compute}
\begin{tabular}{lrr}
\toprule
Experiment & Wall-clock & GPU-hours \\
\midrule
Price-vector calibration & 0.5\,h & 2 \\
Residency characterization & 2.6\,h & 10.3 \\
Capacity and calibration probes & 0.7\,h & 2.6 \\
Live end-to-end serving (15 runs) & 14\,h & 55 \\
Replay and trace generation & CPU only & --- \\
\midrule
Total reported & ${\sim}18$\,h & ${\sim}70$ \\
\bottomrule
\end{tabular}
\end{table}

The full project, including exploratory and invalidated runs not reported in
the paper, consumed approximately 250--300 GPU-hours.

\FloatBarrier

\section{Licenses for external assets}
\label{app:licenses}

We use four primary external research assets. The $\tau^2$-bench code and released benchmark materials are distributed under the MIT License \citep{tau2bench}. vLLM, used as the serving engine in our experiments, is distributed under the Apache License 2.0 \citep{kwon2023vllm}. Continuum, used as an agent-serving baseline, is distributed under the Apache License 2.0 and evaluated using the first author's public fork \citep{continuum_code} (\texttt{vllm-continuum}, commit \texttt{316a587}, 2026-03-20; vLLM v0.10.2 base, no tagged release) as shipped, with its fixed 2\,s pin TTL \citep{continuum2025}. MORI has no public artifact; we evaluate a policy port of its published placement rule on our serving engine \citep{mori2026}. Llama-3.1-70B is provided by Meta under the Llama 3.1 Community License. We use these assets for research and evaluation in accordance with their respective license terms and do not redistribute modified model weights or third-party benchmark data as part of this work.


\end{document}